\documentclass[ALICE,manyauthors]{cernphprep}
\usepackage[comma,square,numbers,sort&compress]{natbib}
\usepackage{hyperref}
\usepackage{lineno}
\usepackage{xspace}
\usepackage{bm}
\usepackage{mathrsfs}
\usepackage{upgreek}
\usepackage[version=4]{mhchem}
\usepackage[T1]{fontenc}
\usepackage{orcidlink}

\begin{document}
%

\newcommand{\pp}           {pp\xspace}
\newcommand{\ppbar}        {\mbox{$\mathrm {p\overline{p}}$}\xspace}
\newcommand{\XeXe}         {\mbox{Xe--Xe}\xspace}
\newcommand{\PbPb}         {\mbox{Pb--Pb}\xspace}
\newcommand{\pA}           {\mbox{pA}\xspace}
\newcommand{\pPb}          {\mbox{p--Pb}\xspace}
\newcommand{\AuAu}         {\mbox{Au--Au}\xspace}
\newcommand{\dAu}          {\mbox{d--Au}\xspace}

\newcommand{\s}            {\ensuremath{\sqrt{s}}\xspace}
\newcommand{\snn}          {\ensuremath{\sqrt{s_{\mathrm{NN}}}}\xspace}
\newcommand{\pt}           {\ensuremath{p_{\rm T}}\xspace}
\newcommand{\kt}           {\ensuremath{k_{\rm T}}\xspace} 
\newcommand{\meanpt}       {$\langle p_{\mathrm{T}}\rangle$\xspace}
\newcommand{\minv}         {\ensuremath{m_{\rm inv}}\xspace}
\newcommand{\mee}          {\ensuremath{m_{\rm e^+e^-}}\xspace}
\newcommand{\ptee}         {\ensuremath{p_{\rm T}^{\rm ee}}\xspace}
\newcommand{\ycms}         {\ensuremath{y_{\rm CMS}}\xspace}
\newcommand{\ylab}         {\ensuremath{y_{\rm lab}}\xspace}
\newcommand{\etarange}[1]  {\mbox{$\left | \eta \right |~<~#1$}}
\newcommand{\yrange}[1]    {\mbox{$\left | y \right |~<~#1$}}
\newcommand{\dndy}         {\ensuremath{\mathrm{d}N_\mathrm{ch}/\mathrm{d}y}\xspace}
\newcommand{\dndeta}       {\ensuremath{\mathrm{d}N_\mathrm{ch}/\mathrm{d}\eta}\xspace}
\newcommand{\avdndeta}     {\ensuremath{\langle\dndeta\rangle}\xspace}
\newcommand{\dNdy}         {\ensuremath{\mathrm{d}N_\mathrm{ch}/\mathrm{d}y}\xspace}
\newcommand{\Npart}        {\ensuremath{N_\mathrm{part}}\xspace}
\newcommand{\Ncoll}        {\ensuremath{N_\mathrm{coll}}\xspace}
\newcommand{\dEdx}         {\ensuremath{\textrm{d}E/\textrm{d}x}\xspace}
\newcommand{\RpPb}         {\ensuremath{R_{\rm pPb}}\xspace}
\newcommand{\Raa}         {\ensuremath{R_{\rm AA}}\xspace}
\newcommand{\mpt}          {\ensuremath{\langle p_{\rm T}\rangle}\xspace}
\newcommand{\mptsquared}   {\ensuremath{\langle p^2_{\rm T}\rangle}\xspace}
\newcommand{\z}            {\ensuremath{z^{\rm{ch}}}\xspace}
\newcommand{\ptjetch}           {\ensuremath{p_{\rm T}^{\text{ch jet}}}\xspace}
\newcommand{\ptjet}           {\ensuremath{p_{\rm T}^{\text{jet}}}\xspace}
\newcommand{\ptjpsi}           {\ensuremath{p_{\rm T}^{\text{J}/\psi}}\xspace}

\newcommand{\onethreeH}       {$\sqrt{s_{\mathrm{NN}}}~=~130$~Ge\kern-.1emV\xspace}
\newcommand{\nineH}        {$\sqrt{s}~=~0.9$~Te\kern-.1emV\xspace}
\newcommand{\oneninesix}  {$\sqrt{s}~=~1.96$~Te\kern-.1emV\xspace}
\newcommand{\twoH}         {$\sqrt{s}~=~200$~Ge\kern-.1emV\xspace}
\newcommand{\twosevensix}  {$\sqrt{s}~=~2.76$~Te\kern-.1emV\xspace}
\newcommand{\five}         {$\sqrt{s}~=~5.02$~Te\kern-.1emV\xspace}
\newcommand{\fivefive}         {$\sqrt{s_{\mathrm{NN}}}~=~5.5$~Te\kern-.1emV\xspace}
\newcommand{\seven}        {$\sqrt{s}~=~7$~Te\kern-.1emV\xspace}
\newcommand{\eight}        {$\sqrt{s}~=~8$~Te\kern-.1emV\xspace}
\newcommand{\thirteen}     {$\sqrt{s}~=~13$~Te\kern-.1emV\xspace}
\newcommand{\thirteensix}     {$\sqrt{s}~=~13.6$~Te\kern-.1emV\xspace}
\newcommand{\fourteen}     {$\sqrt{s}~=~14$~Te\kern-.1emV\xspace}
\newcommand{\twosevensixnn}{$\sqrt{s_{\mathrm{NN}}}~=~2.76$~Te\kern-.1emV\xspace}
\newcommand{\fivenn}       {$\sqrt{s_{\mathrm{NN}}}~=~5.02$~Te\kern-.1emV\xspace}
\newcommand{\LT}           {L{\'e}vy-Tsallis\xspace}
\newcommand{\GeVc}         {Ge\kern-.1emV/$c$\xspace}
\newcommand{\MeVc}         {Me\kern-.1emV/$c$\xspace}
\newcommand{\TeV}          {Te\kern-.1emV\xspace}
\newcommand{\GeV}          {Ge\kern-.1emV\xspace}
\newcommand{\keV}          {ke\kern-.1emV\xspace}
\newcommand{\MeV}          {Me\kern-.1emV\xspace}
\newcommand{\GeVmass}      {Ge\kern-.2emV/$c^2$\xspace}
\newcommand{\MeVmass}      {Me\kern-.2emV/$c^2$\xspace}
\newcommand{\lumi}         {\ensuremath{\mathcal{L}}\xspace}

\newcommand{\ITS}          {\rm{ITS}\xspace}
\newcommand{\TOF}          {\rm{TOF}\xspace}
\newcommand{\ZDC}          {\rm{ZDC}\xspace}
\newcommand{\ZDCs}         {\rm{ZDCs}\xspace}
\newcommand{\ZNA}          {\rm{ZNA}\xspace}
\newcommand{\ZNC}          {\rm{ZNC}\xspace}
\newcommand{\SPD}          {\rm{SPD}\xspace}
\newcommand{\SDD}          {\rm{SDD}\xspace}
\newcommand{\SSD}          {\rm{SSD}\xspace}
\newcommand{\TPC}          {\rm{TPC}\xspace}
\newcommand{\TRD}          {\rm{TRD}\xspace}
\newcommand{\VZERO}        {\rm{V0}\xspace}
\newcommand{\VZEROA}       {\rm{V0A}\xspace}
\newcommand{\VZEROC}       {\rm{V0C}\xspace}
\newcommand{\Vdecay} 	   {\ensuremath{V^{0}}\xspace}

\newcommand{\jpsi}         {\ensuremath{\text{J}/\psi}\xspace}
\newcommand{\psiprime}     {\ensuremath{\psi(2\text{S})}\xspace}
\newcommand{\ccbar}        {\ensuremath{\text{c}\overline{\text{c}}}\xspace}
\newcommand{\bbbar}        {\ensuremath{\text{b}\overline{\text{b}}}\xspace}
\newcommand{\qqbar}        {\ensuremath{\text{q}\overline{\text{q}}}\xspace}
\newcommand{\ee}           {\ensuremath{\text{e}^{+}\text{e}^{-}}\xspace} 
\newcommand{\pizero}          {\ensuremath{\pi^{0}}\xspace}
\newcommand{\pip}          {\ensuremath{\pi^{+}}\xspace}
\newcommand{\pim}          {\ensuremath{\pi^{-}}\xspace}
\newcommand{\kap}          {\ensuremath{\rm{K}^{+}}\xspace}
\newcommand{\kam}          {\ensuremath{\rm{K}^{-}}\xspace}
\newcommand{\pbar}         {\ensuremath{\rm\overline{p}}\xspace}
\newcommand{\kzero}        {\ensuremath{{\rm K}^{0}_{\rm{S}}}\xspace}
\newcommand{\dzero}        {\ensuremath{{\rm D}^{0}}\xspace}
\newcommand{\dpluss}        {\ensuremath{{\rm D}^{+}}\xspace}
\newcommand{\dstar}        {\ensuremath{{\rm D}^{\ast+}}\xspace}
\newcommand{\lmb}          {\ensuremath{\Lambda}\xspace}
\newcommand{\almb}         {\ensuremath{\overline{\Lambda}}\xspace}
\newcommand{\Om}           {\ensuremath{\Omega^-}\xspace}
\newcommand{\Mo}           {\ensuremath{\overline{\Omega}^+}\xspace}
\newcommand{\X}            {\ensuremath{\Xi^-}\xspace}
\newcommand{\Ix}           {\ensuremath{\overline{\Xi}^+}\xspace}
\newcommand{\Xis}          {\ensuremath{\Xi^{\pm}}\xspace}
\newcommand{\Oms}          {\ensuremath{\Omega^{\pm}}\xspace}
\newcommand{\degree}       {\ensuremath{^{\rm o}}\xspace}

\newcommand{\note}[1]{{\color{red}\textbf{#1}}}

\begin{titlepage}
\PHyear{2026}       
\PHnumber{019}      
\PHdate{28 January}  

\title{Measurement of the charged-particle-jet transverse-momentum fraction carried by prompt and non-prompt J/$\bm{\psi}$ mesons in pp collisions at $\sqrt{\bm{s}}$~=~13~TeV}

\ShortTitle{\texorpdfstring{ \unboldmath In-jet J/$\psi$ production in pp collisions at $\s = 13$~TeV}{In-jet J/psi production in pp collisions at sqrt(s) = 13 TeV}}   

\Collaboration{ALICE Collaboration\thanks{See Appendix~\ref{app:collab} for the list of collaboration members}}
\ShortAuthor{ALICE Collaboration} 

\begin{abstract}
The measurement of the transverse-momentum fraction (\z) carried by prompt and non-prompt \jpsi in charged-particle jets in proton--proton collisions with a center-of-mass energy \thirteen is reported by the ALICE Collaboration at the CERN Large Hadron Collider. The measurement is based on a Transition Radiation Detector triggered data sample corresponding to an integrated luminosity of $\mathcal{L} = 1.63\pm0.03 ~\rm pb^{-1}$. Inclusive \jpsi mesons with transverse momentum (\pt) above 1~\GeVc are reconstructed at midrapidity through the electron--positron decay channel. The prompt and non-prompt \jpsi contributions are separated using the secondary vertices from beauty-hadron decays.  Jets are reconstructed from \jpsi-meson candidates and charged particles using the anti-\kt algorithm with jet resolution parameter $R = 0.4$ in the pseudorapidity range $|\eta_{\text{jet}}| < 0.5$. The distribution of the charged-particle jet \pt fraction carried by the prompt and non-prompt \jpsi mesons, \z, is measured in the range $0.3 < \z \leq 1.0$ for the jet-\pt range $7 < \ptjetch < 15$~\GeVc.
The measured \z distributions for prompt and non-prompt \jpsi are compared to pQCD calculations and PYTHIA 8 simulations.
Prompt and non-prompt \jpsi production within charged-particle jets are found to be qualitatively well reproduced by PYTHIA 8 for $\z<0.9$, however, as \z approaches one, the simulations overshoot the measured data, indicating an overestimation of the fraction of isolated \jpsi. This observed tension between data and simulations in the highest \z bins reflects the challenge of simulating hadronization of low-momentum jets. 
\end{abstract}
\end{titlepage}

\setcounter{page}{2} 


\section{Introduction} 
\label{sec:Introduction}
The production of charmonium states, i.e., the creation of bound states of charm--anticharm quark pairs, is a process which cannot be calculated perturbatively in quantum chromodynamics (QCD), and is therefore a valuable testing ground for QCD-based models~\cite{Brambilla:2010cs, ALICE:2022wpn} attempting to understand hadronization. Charmonium-production models usually couple fixed-order perturbative calculations describing the creation of \ccbar pairs with non-perturbative matrix elements describing the evolution into a bound state. Phenomenological approaches, such as the Color Evaporation Model (CEM)~\cite{Amundson:1996qr, Fritzsch:1977ay, Lansberg:2020rft}, the Color Singlet Model (CSM)~\cite{Lansberg:2011hi}, and the non-relativistic QCD (NRQCD) factorization approach~\cite{Butenschoen:2012qr, Chao:2012iv, Ma:2010jj} differ mainly in the description of the charmonium formation process. A set of high-precision cross section and polarization measurements of the lightest charmonium vector state, \jpsi, made at the Large Hadron Collider (LHC)~\cite{LHCb:2012kaz, ALICE:2012vup, ALICE:2019pid,  ALICE:2016flj, ALICE:2017leg, ATLAS:2017prf, CMS:2017exb, CMS:2017uuv, ALICE:2021edd, ALICE:2021qlw, LHCb:2021pyk, ALICE:2012vpz, ALICE:2014uja, LHCb:2011zfl, CMS:2010nis, ATLAS:2011aqv, CMS:2015lbl, ALICE:2015pgg, LHCb:2013itw, ATLAS:2015zdw, ALICE:2021dtt, LHCb:2015foc, CMS:2017dju, ATLAS:2023qnh, ALICE:2011gej, CMS:2013gbz, LHCb:2013izl, ALICE:2018crw, ALICE:2024yqd} have enabled improvements to theoretical descriptions, such as the improved CEM~\cite{Ma:2016exq, Cheung:2018tvq} and several NRQCD-based approaches~\cite{Butenschoen:2012px, Gong:2012ug, Butenschoen:2011yh, Bodwin:2014gia, Faccioli:2014cqa, Bodwin:2015iua, Faccioli:2017hym, Faccioli:2018uik}.
Nevertheless, a universal and successful model of the \jpsi-production mechanism remains elusive, especially in the low transverse-momentum (\pt) range~\cite{ALICE:2021qlw, ALICE:2021dtt, ALICE:2018crw}. 

New observables that correlate charmonium production with hadronic activity in the same event, such as the recent measurement of the hadron yields associated with \jpsi production reported by ALICE in Ref.~\cite{ALICE:2024yqd}, could help improve our understanding of the formation of the bound state.
Studies of \jpsi production in jets -- collimated sprays of hadrons resulting from the fragmentation of high-\pt partons -- may provide further insight into the production phenomenology. Observables, such as the fraction of jet transverse momentum carried by a \jpsi, are expected to be sensitive to both the perturbative and the non-perturbative aspects of its formation~\cite{Baumgart:2014upa}. In particular, the jets where the \jpsi mesons carry most of the jet momentum (isolated \jpsi) are expected to be sensitive to the relative contribution of the hadronization mechanisms via color-singlet and color-octet states leading to \jpsi production.

Measurements of prompt-\jpsi production in jets ($\ptjet > 20$~\GeVc) reported by LHCb~\cite{LHCb:2017llq} at forward rapidity in proton--proton (pp) collisions at \thirteen show that the \pt fraction carried by the \jpsi is overestimated by calculations implemented in PYTHIA 8 using the NRQCD framework. Here, prompt \jpsi refers to directly produced \jpsi mesons and feed-down contributions from higher-mass charmonium states. A similar observation was reported by CMS~\cite{CMS:2021puf} at midrapidity in pp collisions at \five for jets with $30 <\ptjet <40$~\GeVc. 
Likewise, the production of prompt \psiprime and $\chi_{c1}(3872)$ in jets reported by LHCb~\cite{LHCb:2024ybz} at forward rapidity in pp collisions at \thirteen shows striking differences from PYTHIA 8 predictions for $\ptjet >15$~\GeVc. In this jet-\pt range, the measured distributions indicate two distinct components corresponding to non-isolated and isolated production.
The prompt-charmonium production simulated in PYTHIA 8 originates directly from parton--parton interactions. The observed discrepancies may be better described by advanced calculations of charmonium fragmentation functions~\cite{Bain:2016rrv, Bain:2017wvk} and the inclusion of a parton shower prior to the \jpsi formation in PYTHIA 8~\cite{Cooke:2023ldt}.

In the high jet-\pt regime, the jet production cross section can be determined theoretically using the factorization theorem~\cite{Collins:1985ue}. In this limit, parton fragmentation is assumed to be universal, i.e., independent of the collision process. However, in the low-\pt range, the factorization may be broken. Measurements of the charm-quark fragmentation fractions show strong deviations from values measured in \ee and ep collisions~\cite{ALICE:2021dhb}. Studies of identified particles produced within the jet cone may provide new insights into the hypothesis of universality of parton fragmentation. 
Recent measurements of charm baryon-tagged jets within a relatively low \pt region ($7< \ptjetch < 15$~\GeVc) hint at a softer fragmentation into charm baryons with respect to charm mesons~\cite{ALICE:2023jgm}. Investigating the fragmentation of \jpsi in jets in a low jet-\pt region may provide complementary input about the fragmentation of heavy quarks (charm and beauty). 

The study of in-jet charmonium production in \pp collisions is vital for the interpretation of future similar measurements in heavy-ion collisions. At high temperature and close to zero baryon chemical potential, lattice QCD calculations predict a deconfined state of strongly-interacting matter known as the quark--gluon plasma (QGP)~\cite{Bazavov:2009zn, Borsanyi:2013bia}. The charmonium formation process in heavy-ion collisions is complex, with the initial production of heavy-quark pairs predominantly taking place very early in the collision history. The production of bound charmonium states is expected to be strongly affected by the hot medium created in the collision, as the heavy-quark pairs may be subject to effects such as color screening,  (re)generation, and parton energy loss~\cite{Matsui:1986dk, Thews:2000rj,Braun-Munzinger:2000csl, Bjorken:141477}. In particular, the production of charmonia in ultra-relativistic heavy-ion collisions is one of the most prominent probes of the QGP~\cite{Matsui:1986dk}, where phenomenological models differ greatly in the assumed underlying production mechanisms, and accordingly, predicted differential yields. Statistical hadronization models assume that the charm quarks produced in initial hard scatterings reach a complete thermal equilibration in the QGP, and all the bound charmed hadrons are produced at the chemical freeze-out~\cite{Braun-Munzinger:2000csl, Andronic:2019wva}. On the other hand, phenomenological transport models assume that charm hadrons are continuously produced, dissociated, and (re)generated during the entire evolution of the hot and dense matter in the collision zone.~\cite{Zhao:2007hh, Zhou:2014kka}. So far, the various models provide
a relatively good description of nuclear suppression and azimuthal anisotropy (flow) measurements~\cite{ALICE:2023gco, ALICE:2020pvw, He:2021zej}, requiring further measurements to distinguish between the assumed underlying production mechanisms.
Studies of charmonium-tagged jets and the modification of the jet substructure in heavy-ion collisions may provide a complementary insight into charmonium production, the energy-loss mechanisms of heavy quarks, and the properties of the produced deconfined medium~\cite{Caucal:2018dla, Chung:2018lyq}.

This paper presents measurements of \jpsi production within  charged-particle jets (i.e., jets reconstructed using only charged-particle tracks) in pp collisions at a center-of-mass energy of \thirteen. The measurement is performed using a Transition Radiation Detector (TRD) triggered data sample recorded by the ALICE detector at the LHC, corresponding to an integrated luminosity of $\mathcal{L} = 1.6\pm0.03 ~\rm pb^{-1}$.
The fraction of the jet transverse momentum carried by the \jpsi, defined as
\begin{equation}
    \z \equiv \frac{\ptjpsi}{\ptjetch} \, ,
\end{equation}
is determined separately for prompt and non-prompt \jpsi, where the latter refers to \jpsi mesons originating from the decay of beauty hadrons. Jets are reconstructed using the anti-\kt algorithm~\cite{Cacciari:2008gp} with a jet resolution parameter of $R=0.4$. Reconstructed jets are required to be in the pseudorapidity range $|\eta_{\text{jet}}|<0.5$ and have a \pt in the range $7 < \ptjetch < 15$~\GeVc. The reconstruction of \jpsi mesons is done at midrapidity through the electron--positron decay channel for \jpsi with $\ptjpsi>1$~\GeVc and the separation between the prompt and the non-prompt components using the displacement between the primary collision vertex and the \jpsi decay vertex.

\section{Detector setup and data samples}
\label{sec:dataSample}

A full description of the ALICE setup and detector performance can be found in Refs.~\cite{ALICE:2008ngc, ALICE:2014sbx}. In the following, the detectors employed in this analysis and their roles are described. The Inner Tracking System (ITS) is a six-layer silicon detector located around the beam pipe. The two innermost layers of the ITS barrel are Silicon Pixel Detectors (SPD) with excellent spatial resolution placed at a radial distance of $r = 3.9$ and 7.6~cm from the beam line. This detector is used for primary and secondary-vertex reconstruction, and also for tracking. In particular, the high spatial resolution of the ITS enables the measurement of displaced decay vertices, therefore allowing for the separation between the prompt and non-prompt \jpsi.  
The Time Projection Chamber (TPC)~\cite{ALICE:2000jwd} is a cylindrical gaseous detector covering a radial distance between 85 and 250~cm. It is located around the ITS and extends 250~cm in the longitudinal direction on each side of the nominal interaction point. The TPC is used for tracking and particle identification (PID) based on the specific energy loss of particles interacting with the TPC gas. 
The TPC and ITS cover the full azimuthal angle and provide tracking and PID performance in the pseudorapidity range $|\eta|<0.9$. 
The Transition Radiation Detector (TRD)~\cite{ALICE:2017ymw} surrounds the TPC and covers the full azimuthal angle and the pseudorapidity range $|\eta|<0.84$. The TRD consists of 522 chambers arranged in 18 sectors in azimuth and in six radial layers and five stacks in the longitudinal direction. Each chamber consists of a fiber mat sandwiched between two foam sheets operating as radiator material, where highly boosted particles emit transition radiation. 
Behind the radiators, multi-wire proportional drift chambers with pad readout, filled with a \ce{Xe}--\ce{CO2} gas mixture, provide tracking.
In addition, the signals from the drift region provide measurements of the specific energy loss of charged particles and the energy of transition-radiation photons originating from the relativistic electrons crossing the radiator material. Since the tracking and PID signal can be computed online, the TRD is used as an electron trigger, with a decision being issued within 6~$\upmu$s after the collision time. 
Two scintillator arrays (V0 detectors)~\cite{ALICE:2013axi} situated along the beam direction at $-3.7 <\eta < -1.7$ and $2.8 < \eta < 5.1$ are used for the minimum-bias triggering and event characterization.

The analysis discussed in this article is based on data recorded in the years 2017 and 2018 in pp collisions with center-of-mass energy \thirteen. The data sample corresponds to events triggered by the TRD using a condition that at least one TRD track, i.e., a track segment measured in the TRD and constrained to the primary vertex, with a particle \pt higher than 2~\GeVc and a PID signal compatible with the electron hypothesis.
To reduce the contribution to electron-triggered events due to photon conversions in the detector material, the trigger condition also requires the track to have five tracklets per stack, including one in the first TRD layer.
The background from secondary electrons is further reduced using a selection on the pointing angle of the track~\cite{ALICE:2022zig}.
Besides the TRD-triggered data sample, a minimum-bias (MB) data sample is considered to evaluate the TRD-trigger performance in data. MB events are selected based on the coincidence of the signals in the V0 scintillator arrays.

Selected events are required to have a reconstructed primary vertex located within $\pm$ 10~cm of the nominal interaction point along the beam direction. With a maximum interaction rate of 260~kHz, the probability for pileup events from the same bunch crossing is $0.5\times10^{-3}$~\cite{ALICE:2021dtt}. An offline algorithm which identifies multiple vertices~\cite{ALICE:2014sbx} is used to reject these events. Beam--gas events and collisions from other LHC bunch crossings occurring within the readout time of the SPD are rejected based on offline timing cuts using information from the V0.

A simulated Monte Carlo (MC) sample is used in the determination of probability density functions (PDFs) required for the extraction of the non-prompt \jpsi fraction (described in Sec.~\ref{sec:yields}) and for necessary corrections (described in Sec.~\ref{sec:corrections}). The MC sample consists of MB pp events generated with PYTHIA 6.4~\cite{Sjostrand:2006za}, where additional prompt and non-prompt \jpsi mesons are injected. The non-prompt \jpsi mesons originate from the decay of beauty hadrons as implemented in PYTHIA 6, while the prompt \jpsi mesons are simulated with a \pt spectrum based on a phenomenological interpolation of existing cross section measurements~\cite{Bossu:2011qe}. The \jpsi-meson decay is simulated with EvtGen~\cite{Lange:2001uf} using the PHOTOS model~\cite{Barberio:1993qi}, which takes into account the QED radiation at the decay vertex. The generated particles are transported through a simulated ALICE detector setup using GEANT 3~\cite{Brun:1082634}, and then reconstructed using the same workflow as for real data. 
In particular, the TRD trigger is also emulated in the MC simulation by applying the same selection criteria as in data. 
An overview of the number of selected events and corresponding integrated luminosity for the analyzed data and MC samples is shown in Tab.~\ref{tab:dataSampleStat}. 

\begin{table}[tb]
\centering
\caption{Overview of the number of selected events and corresponding integrated luminosity in the analyzed data and MC samples.}
	\begin{tabular}{l |c c | c }
		\hline
        &\multicolumn{2}{c|}{\textit{data}} &\textit{simulation} \\
		Trigger & MB 	& TRD & MC, \jpsi injected \\
		\hline
    Number of events    &$1.06\times 10^{9}$ &$1.71\times 10^{8}$ &$7.86\times 10^{7}$ \\
        Integrated luminosity  &$17.83 \pm 0.33~\rm{nb}^{-1}$  &$1.63 \pm 0.03~\rm{pb}^{-1}$ & -- \\ 
		\hline
	\end{tabular}
	\label{tab:dataSampleStat}
\end{table}

\section{Analysis}
The \jpsi mesons are reconstructed at midrapidity through their electron--positron decay channel. The separation of prompt and non-prompt \jpsi is done on a statistical basis by using the displacement of the reconstructed secondary vertex of the decaying beauty hadron. The jets are reconstructed from charged particles in the central barrel using the anti-\kt algorithm~\cite{Cacciari:2008gp}. The following section describes the analysis in detail.

\subsection{Track selection}
\label{sec:trackSel}
The reconstruction of \jpsi mesons is done by combining electron and positron tracks. The tracks are reconstructed in the central barrel based on information from the ITS and TPC, following the selections used in previous analyses reported in Refs.~\cite{ALICE:2021edd, ALICE:2022zig}. Selected tracks are required to be within the TRD acceptance ($|\eta| < 0.84$) with a minimum \pt of 1~\GeVc. In order to reach a good spatial resolution, needed for the secondary vertexing and to reject background from photon conversions, tracks are required to have at least one hit in the SPD layers. To ensure a good momentum and PID resolution, a minimum of 70 space points in the TPC is also required. In order to further reject secondary electrons, candidates are required to have a maximum distance-of-closest approach (DCA) to the reconstructed collision vertex of 1~cm in the radial direction and 3~cm along the beam axis. Electrons are identified by their specific energy loss measured in the TPC, and are selected to be within a band of $\pm 3\sigma$ relative to the expected electron energy-loss curve, where $\sigma$ represents the resolution of the energy loss associated with each track. To minimize contamination from protons and pions, electron-candidate tracks compatible with the proton or pion energy loss expectation within $3\sigma$ are rejected. In order to increase the \jpsi-reconstruction efficiency at high \pt, a rejection criteria of $2\sigma$ is chosen for electron candidates with momentum above 5~\GeVc. 

For the jet reconstruction, a less stringent track selection is applied in order to provide a uniform acceptance in $\eta$ and $\varphi$. Charged-particle tracks are required to be within the acceptance of the central barrel ($|\eta| < 0.9$) with a minimum \pt of 0.15~\GeVc. Jets are reconstructed using the anti-\kt algorithm~\cite{Cacciari:2008gp} as implemented in the FastJet package~\cite{Cacciari:2011ma} using the \pt scheme for recombination. The jet reconstruction is performed with a jet resolution parameter $R = 0.4$ and a minimum jet \ptjetch of 5~\GeVc. Jets are required to have a pseudorapidity of $|\eta_{\text{jet}}| < 0.5$ to ensure that they stay fully within the detector acceptance.
The possible bias introduced by \jpsi decay electrons restricted to the TRD-trigger acceptance while jets are restricted by the central-barrel acceptance was found to be negligible. Jet reconstruction is performed in events where at least one \jpsi candidate is found, where the decay electrons corresponding to a \jpsi candidate are removed before the jet reconstruction and are replaced by the four-momentum of the reconstructed \jpsi candidate. In the rare case of more than one \jpsi candidate in an event, this step is done separately for each \jpsi candidate. After the jet-reconstruction procedure, the fraction of \pt carried by the \jpsi candidate within its jet, \z, is obtained. No subtraction of the underlying event is applied, since the background in this \pt range is negligible for pp collisions.

\subsection{\texorpdfstring{Extraction of the prompt and non-prompt J/$\psi$ yields}{Extraction of the prompt and non-prompt J/psi yields}}
\label{sec:yields}
The extraction of the \jpsi yields in each \z interval relies on building the invariant mass of all possible electron--positron pairs from the same event. Additionally, for the separation of the non-prompt \jpsi component, originating from beauty-hadron decays, the secondary vertex for each pair is also constructed. The method consists of a two-dimensional likelihood fit of the invariant-mass (\mee) and pseudo-proper decay-length distribution ($x$). The pseudo-proper decay length is defined as
\begin{equation}
	x = c \times L_{xy}^{\jpsi}\times \frac{M^{\jpsi}}{\ptjpsi} \, ,
\end{equation}
where $L_{xy}^{\jpsi}$ represents the transverse projection of the distance between the primary and the secondary vertices, $M^{\jpsi}$ is the world-average value of the \jpsi mass~\cite{ParticleDataGroup:2022pth}, and \ptjpsi is the transverse momentum of the \jpsi candidate. 

The fit procedure maximizes the logarithm of a likelihood function
\begin{equation}
 \ln  \{ \mathscr{L} \} = \sum_{i=1}^{N} \ln \{ f_{\text{Sig}} \times \mathcal{F}_{\text{Sig}}(x) \times \mathcal{M}_{\text{Sig}}(\mee) + (1-f_{\text{Sig}})\times \mathcal{F}_{\text{Bkg}}(x) \times \mathcal{M}_{\text{Bkg}}(\mee)  \} \, ,
\end{equation}
where $N$ is the total number of \jpsi candidates in the invariant-mass range $2.4 < \mee < 3.6$~\GeVmass. The terms $\mathcal{F}_{\text{Sig}}(x)$ and $\mathcal{F}_{\text{Bkg}}(x)$ represent the probability density functions (PDFs) of the pseudo-proper decay-length distributions for signal and background, respectively. In the same manner, the terms $\mathcal{M}_{\text{Sig}}(\mee)$ and $\mathcal{M}_{\text{Bkg}}(\mee)$ represent the PDFs describing the invariant-mass distributions for signal and background, respectively.  
The fraction of \jpsi signal within the invariant-mass range defined above is represented by $f_{\text{Sig}}$ and is a free parameter of the fit. The PDF describing the pseudo-proper decay-length distribution of the \jpsi signal includes two components
\begin{equation}
	\mathcal{F}_{\text{Sig}}(x) = f'_{\text{B}} \times \mathcal{F}_{\text{B}}(x) + (1-f'_{\text{B}}) \times \mathcal{F}_{\text{prompt}}(x) \, ,
\end{equation}
where $\mathcal{F}_{\text{prompt}}(x)$ and $\mathcal{F}_{\text{B}}(x)$ correspond to the prompt and non-prompt \jpsi, respectively. The term $f'_{\text{B}}$ represents the uncorrected fraction of non-prompt \jpsi and is also a free parameter of the fit. A complete description of the different PDFs entering the maximum-likelihood fit is given in Refs.~\cite{ALICE:2012vpz, ALICE:2015nvt}. In this analysis, the PDFs corresponding to the \jpsi signal, $\mathcal{M}_{\text{Sig}}(\mee)$, $\mathcal{F}_{\text{prompt}}(x)$, and $\mathcal{F}_{\text{B}}(x)$, are obtained based on a MC simulation. Since both the invariant mass and the $x$ PDFs are \pt dependent, a weighted average of each of these PDFs is employed for every considered \z interval, according to 
\begin{equation}
    T(\z) = \frac{\sum_{\pt} T(\ptjpsi) \times N_{\ee}(\ptjpsi;\z)}{N_{\ee}(\z)} ,
    \label{eq:weightProcedure}
\end{equation}
where $T(\ptjpsi)$ is a \jpsi signal PDF in a given \ptjpsi interval, and $N_{\ee}(\ptjpsi;\z)$ is the number of electron--positron pairs with transverse mometum \ptjpsi in the \z interval of interest.

The PDFs describing the background distributions, $\mathcal{M}_{\text{Bkg}}(\mee)$ and $\mathcal{F}_{\text{Bkg}}(x)$ are determined from data. The invariant-mass distribution of the background is parametrized by a third-order polynomial function for $\z<0.7$ and an exponential for $\z >0.7$.
The PDF describing the pseudo-proper decay-length distribution of the background is determined by fitting the distributions in the sideband regions $2.4 < \mee < 2.6$~\GeVmass and $3.2 < \mee < 3.6$~\GeVmass with a functional form described in detail in Ref.~\cite{ALICE:2012vpz}.

\begin{figure}[tb]
    \begin{center}
    \includegraphics[width = 0.8\textwidth]{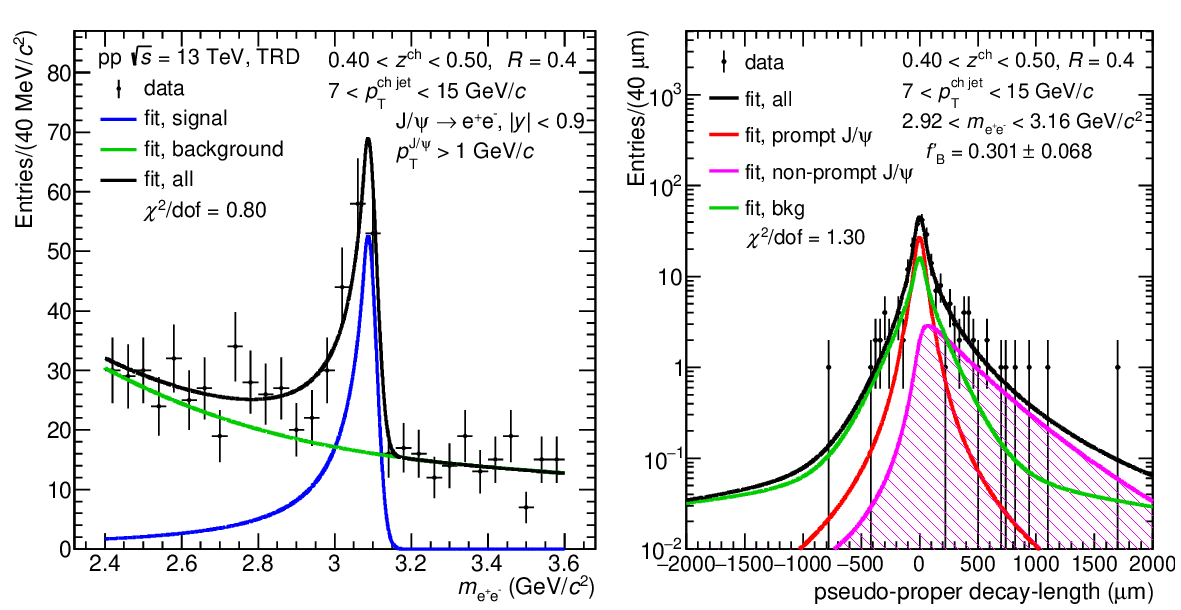}
    \includegraphics[width = 0.8\textwidth]{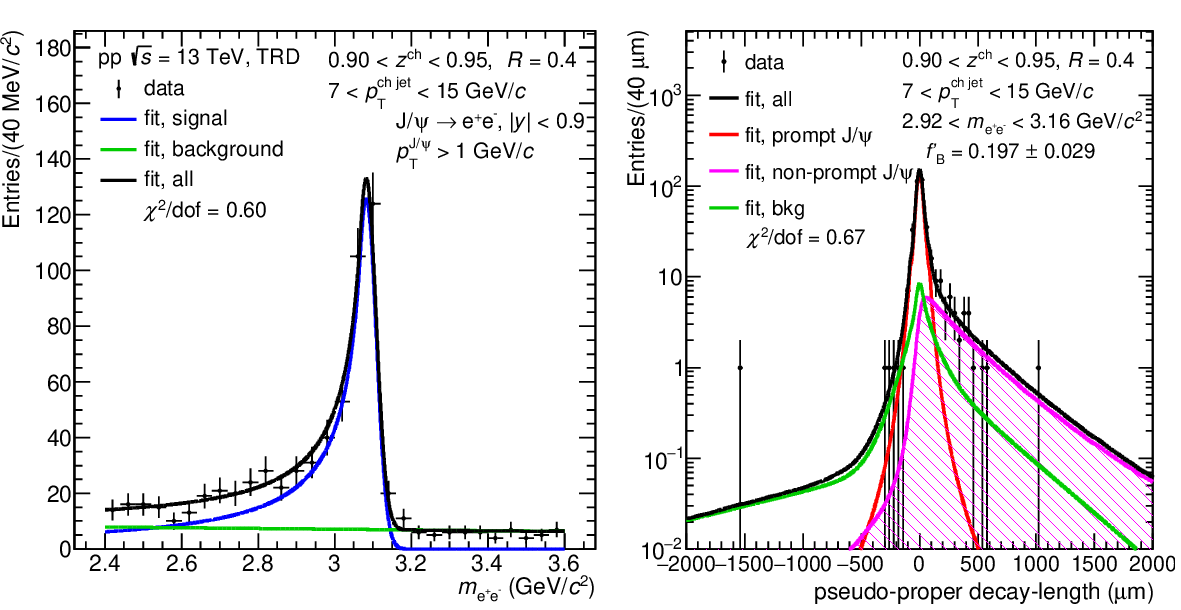}
    \end{center}
    \caption{Invariant-mass (\textit{left}) and pseudo-proper decay-length (\textit{right}) distributions for \jpsi candidates, measured in charged-particle jets with $7<\ptjetch<15$~\GeVc, with superimposed projections of the maximum-likelihood fit for the intervals $0.4 < \z < 0.5$ and $0.90 < \z < 0.95$ in the top and bottom panels, respectively. The pseudo-proper decay-length distributions are shown for \jpsi candidates reconstructed in the invariant-mass range $2.92 < \mee < 3.16$~\GeVmass for display purposes only.}
    \label{fig:likelihoodFit}
\end{figure}

Superimposed projections of the total maximum-likelihood fit function on the invariant-mass and pseudo-proper decay-length distributions are shown in Fig.~\ref{fig:likelihoodFit} for two different \z bins. The maximum-likelihood fit is done in bins of \z for three different jet \ptjetch ranges, further described in section~\ref{sec:unfolding}.

\subsection{Acceptance and efficiency corrections}
\label{sec:corrections}

\begin{figure}[tb]
    \begin{center}
    \includegraphics[width = 0.6\textwidth]{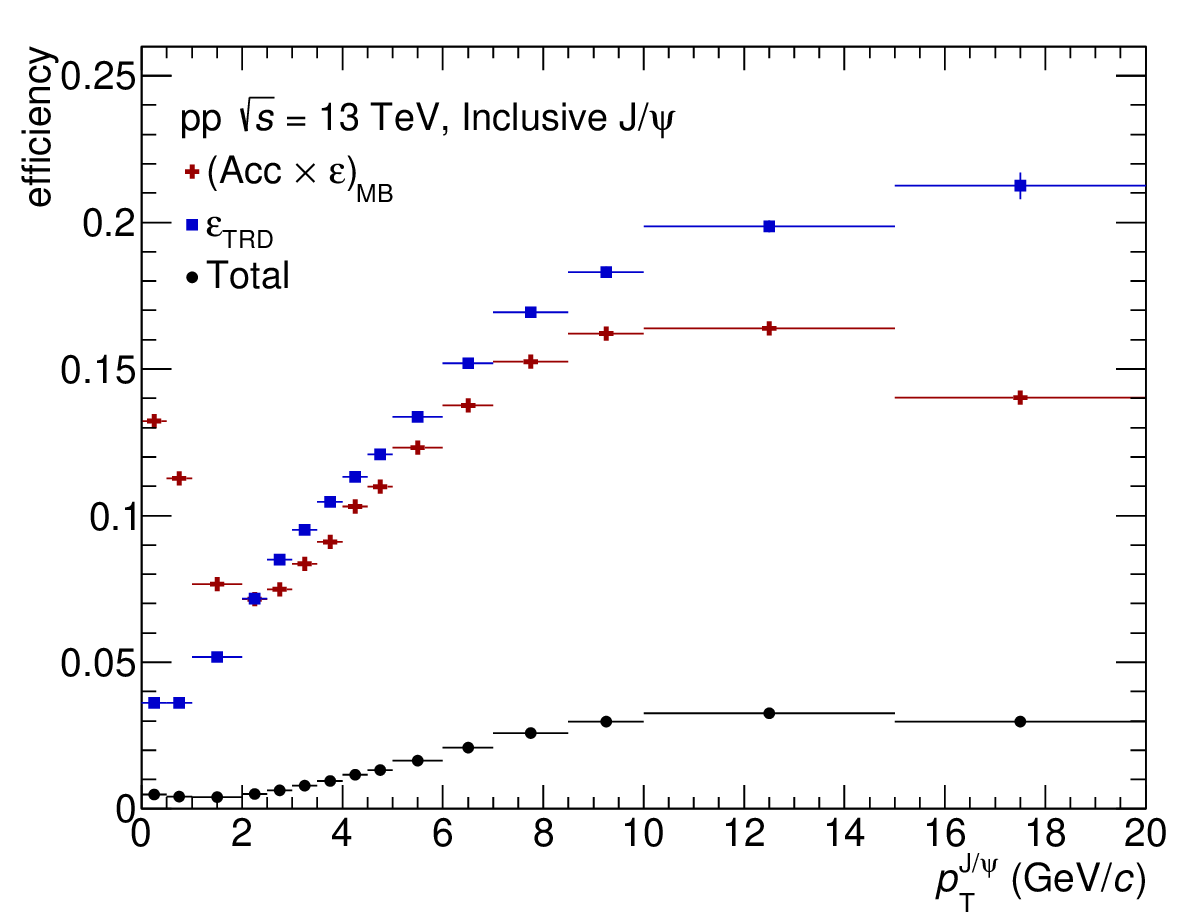}
    \end{center}
    \caption{\jpsi acceptance and reconstruction efficiency (red cross), TRD trigger efficiency (blue box), and total efficiency (black point) as a function of transverse momentum.}
    \label{fig:efficiency}
\end{figure}

For a given \z interval, the corrected yields of prompt ($N_{\text{prompt}}$) and non-prompt ($N_{\text{non-prompt}}$) \jpsi are obtained as 
\begin{align}
    N_{\text{prompt}} &= (1-f_{\text{B}}) \times N_{\text{inclusive}}  \, , \\
    N_{\text{non-prompt}} &= f_{\text{B}} \times N_{\text{inclusive}} \, ,
\end{align}
where $N_{\text{inclusive}}$ and $f_{\text{B}}$ are the corrected number of inclusive \jpsi and the corrected fraction of non-prompt \jpsi, respectively. The corrected number of inclusive \jpsi is obtained as
\begin{equation}
    N_{\rm inclusive} = \frac{N_{\rm inclusive}^{\rm raw}}{(\mathcal{A} \times \varepsilon) \times \varepsilon_{\rm TRD}} ,
    \label{eq:corrYield}
\end{equation}
where $N_{\rm inclusive}^{\rm raw}$ is the number of raw \jpsi-signal counts obtained by summing the invariant-mass distribution after the subtraction of the background distribution resulting from the fit. The $(\mathcal{A} \times \varepsilon)$ is a correction factor obtained from simulations, accounting for the \jpsi kinematic acceptance, the reconstruction efficiency, and the fraction of \jpsi signal found within the invariant-mass window used to count the signal. The TRD trigger efficiency, $\varepsilon_{\rm TRD}$, is determined from simulations following the procedure outlined in Ref.~\cite{ALICE:2022zig}. The algorithm for the TRD trigger is emulated in the MC simulations, where quantities such as the electron likelihood, the online reconstruction of \pt, and the number of track segments per stack are calculated to derive an analogous trigger decision. Figure~\ref{fig:efficiency} shows the total acceptance and efficiency, and the corresponding efficiency contributions. 

Although for a given \ptjpsi value, the acceptance times efficiency is the same for prompt and non-prompt \jpsi, when considering finite \ptjpsi intervals, the average correction factors are slightly different due to the \ptjpsi dependence of the $(\mathcal{A} \times \varepsilon)$ and the fact that the prompt and non-prompt \jpsi have different \ptjpsi distributions. This is taken into account by applying a correction to the raw $f'_{\text{B}}$ obtained from the fit to the data, such that the corrected $f_{\text{B}}$ is obtained as
\begin{equation}
	f_{\text{B}} = \Big( 1 + \frac{1-f'_{\text{B}}}{f'_{\text{B}}} \times \frac{\langle \mathcal{A} \times \varepsilon \rangle_{\text{B}}}{\langle \mathcal{A} \times \varepsilon \rangle_{\text{prompt}}} \Big)^{-1} \, ,
\end{equation}
where $f'_{\text{B}}$ is obtained from the fit, and $\langle \mathcal{A} \times \varepsilon \rangle_{\text{prompt}}$ and $\langle \mathcal{A} \times \varepsilon \rangle_{\text{B}}$ represent the weighted average acceptance and efficiency of the prompt and non-prompt \jpsi in a given \z range, respectively, obtained following the same weighting procedure as described by Eq.~\ref{eq:weightProcedure}. The efficiency-corrected $f_{\text{B}}$ values are in the range 0.15--0.41 for $7<\ptjetch < 15$~\GeVc.

\subsection{Unfolding}
\label{sec:unfolding}
\begin{figure}[tb]
    \begin{center}
    \includegraphics[width = 0.8\textwidth]{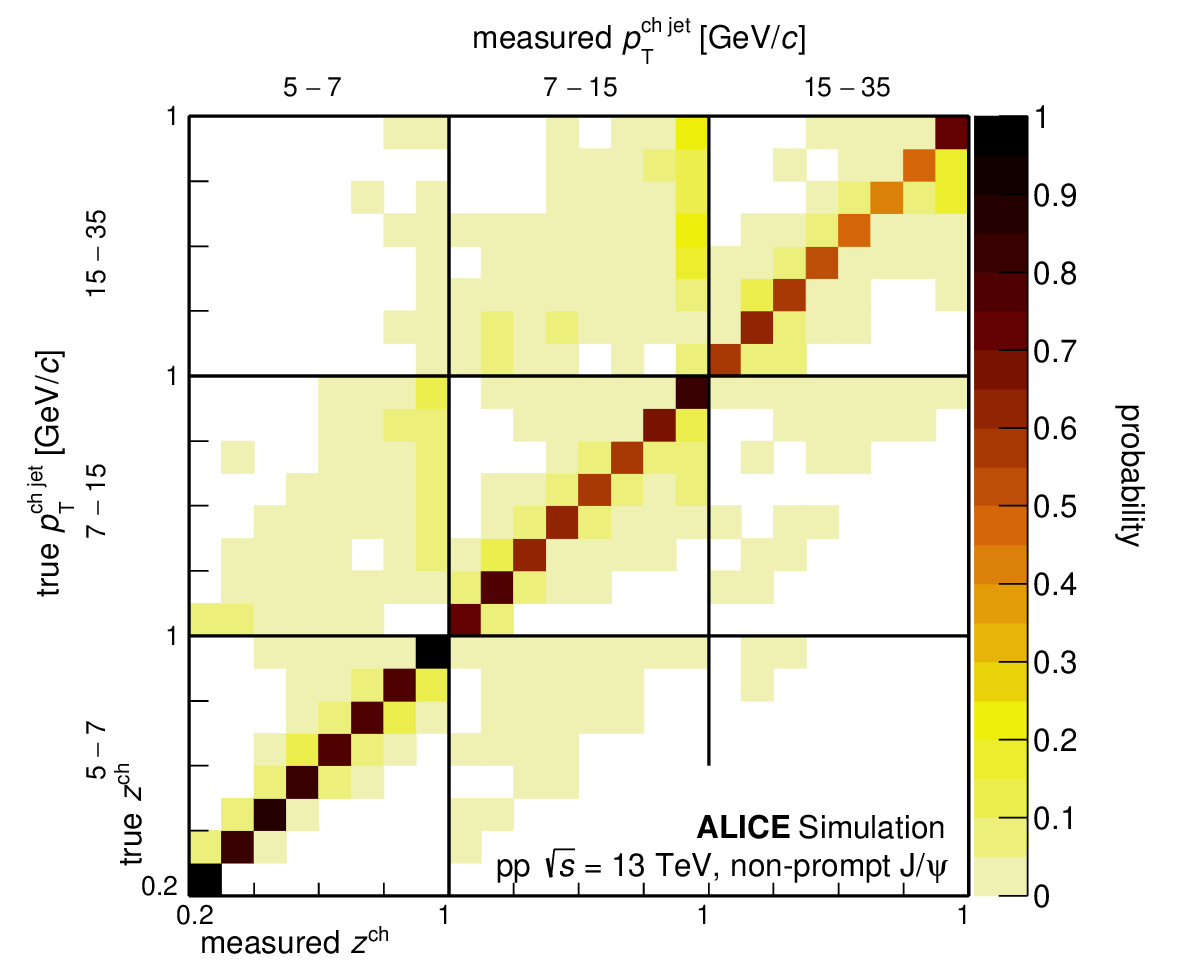}
    \end{center}
    \caption{Detector-response matrix for jets containing a non-prompt \jpsi meson, representing the bin-migration probability as a function of the jet \ptjetch and \z. The $y$ axis represents the true \z in bins of true \ptjetch, and the $x$ axis shows the measured \z in bins of measured \ptjetch.}
    \label{fig:responseMatrix}
\end{figure}

A two-dimensional (\z, \ptjetch) Bayesian unfolding procedure~\cite{DAGOSTINI1995487}, as implemented in the RooUnfold package~\cite{rooUnfold21}, is applied to correct for bin migration due to detector resolution and inefficiencies. The correction is encoded in a 4-dimensional response matrix, which maps the generator-level variables (\z, \ptjetch) to the corresponding detector-level variables. Here, generator-level jets correspond to MC-truth-level charged particles passing the kinematic selections $|\eta| < 0.9$ and $\pt > 0.15$~\GeVc reconstructed using the jet finding procedure described in Sec.~\ref{sec:trackSel}. The response matrix is populated with generator- and detector-level \jpsi-tagged jets obtained from the MC sample described in Sec.~\ref{sec:dataSample}. The generator- and detector-level jets are matched by requiring the same \jpsi candidate among the jet constituents. Due to the significantly larger amount of computing required to generate jets with prompt \jpsi with the ALICE Run 2 MC framework, the response matrix is constructed using only non-prompt \jpsi-tagged jets and used to unfold both the prompt and the non-prompt \jpsi \z distributions. In \pp collisions, the charged-track reconstruction performance in the central barrel is independent of the track density and the measurements are therefore not expected to be affected by this choice. Figure~\ref{fig:responseMatrix} shows the projected response matrix, representing the probability of bin migration, where each row of generated bins is normalized to unity. 

The unfolded results are reported in the $7< \ptjetch < 15$~\GeVc and $0.3 < \z \leq 1.0$ range. In order to account for migrations below and above these intervals, the response matrix is constructed to cover the \pt interval $5 < \ptjetch < 35$~\GeVc and the \z interval $0.2 < \z \leq 1.0$. Additional residual corrections are applied to account for jets migrating in and out of the response matrix range. The corrections are found to be minor.

Several checks are performed to evaluate the stability of the unfolding procedure. Convergence of the unfolding procedure is tested by comparing results of subsequent Bayesian iterations, where convergence is reached after five iterations. This is chosen as the default number of iterations. The unfolding procedure is further tested by refolding the unfolded results and comparing to the measured data. Additionally, a closure test is performed, where the MC sample is split into two independent samples containing 20 and 80\% of the statistics. The first sample is used to provide the input distribution for the unfolding procedure, while the second sample is used to construct the response matrix. The closure test assesses how accurately the unfolding of the input spectrum recreates the original MC-generated spectrum.
The closure test is performed twice, once with statistical uncertainties on the input spectra provided from the MC distribution and once with statistical uncertainties propagated from the measured distribution in data. In both cases, the ratio between the unfolded and generated distributions fluctuates within a 10\% band, with the ratio being statistically compatible with one. 

\section{Systematic uncertainties}
The measurement of \jpsi yields in jet \z intervals is affected by several independent sources of systematic uncertainties due to electron-candidate selection, \jpsi-signal extraction, separation of prompt and non-prompt \jpsi, TRD trigger efficiency, track-reconstruction efficiency, and unfolding. The evaluation of the different systematic sources follows closely the procedures outlined in Refs.~\cite{ALICE:2024yqd, ALICE:2021edd, ALICE:2022zig, ALICE:2023jgm}.

The systematic uncertainty related to the identification of electron tracks is evaluated by varying the PID selection criteria. Three different PID criteria are investigated, namely a tight and loose hadron (protons and pions combined) rejection and a tighter electron selection criterion. For each of these variations, the analysis is repeated and the RMS of the considered variations is assigned as the systematic uncertainty. The uncertainty is considered as correlated between \z bins, and ranges between 1--2\% apart from the lowest \z bin where the uncertainty is about 13\%. 

The uncertainty on the \jpsi-signal extraction has contributions from both the signal and background fit components. The uncertainty related to the signal component is evaluated by changing the invariant-mass interval in which the \jpsi yield is counted and consequently the correction on the fraction of the yield being counted. Five variations are considered, changing the lower bound by $-80$~\MeVmass and the upper bound by $\pm 40$~\MeVmass. The uncertainty on the signal component is considered as uncorrelated between \z bins and ranges from 1--8\%, where the uncertainty is largest for the lowest \z bin.
The systematic uncertainty related to the background PDF of the invariant mass is determined by varying the fit function to either a second-order polynomial or an exponential (third-order polynomial for $\z >0.7$). In addition, the upper and lower bounds of the background fit range are varied by $\pm 100$~\MeVmass. The RMS of the variations of the fit function and fit range is assigned as the systematic uncertainty, and ranges from 1--3\%.

The systematic uncertainty of the TRD trigger efficiency is evaluated by varying the threshold of the electron and positron likelihood value for the trigger decision. The variation corresponds to the granularity with which the electron/positron likelihood is calculated in the front-end electronics. For each variation, the TRD trigger efficiency is recalculated and used to compute the \z-dependent \jpsi yields. The RMS of the deviation from the default case is found to be less than 1\%. The systematic uncertainty of the TRD trigger efficiency is treated as correlated between the \z bins.

The systematic uncertainty on the separated prompt and non-prompt \jpsi yields is determined from the uncertainty on the pseudo-proper decay-length PDFs for the background, prompt, and non-prompt \jpsi.
The procedure to extract these uncertainties is similar to that performed in previous pp analyses reported in Refs.~\cite{ALICE:2021edd, ALICE:2012vpz}, which includes a plausible modification of the resolution function, changes in the non-prompt input template, and a change in the computation of the primary vertex. In addition, the impact of the input spectrum's \pt shape for prompt \jpsi in the MC is evaluated.  The uncertainties related to the prompt and non-prompt pseudo-proper decay-length PDFs are considered to be correlated and range from 1--6\%, where the uncertainty decreases for increasing \z bins. 
 
The systematic uncertainty related to the background fit of the pseudo-proper decay-length distribution is determined by varying the upper and lower sideband regions in steps of $\pm 50$ and $\pm 100$~\MeVmass.
The RMS of the considered variations is assigned as the systematic uncertainty related to the pseudo-proper decay-length background PDF. The uncertainty is considered as uncorrelated between the different \z intervals and ranges from 1--6\%, apart from the lowest \z bin where the uncertainty is 11\%. 

The systematic-uncertainty contribution from unfolding is estimated by varying the default MC generator-level prior distribution to a flat \z distribution. Additionally, the default number of five iterations is varied by $\pm 1$. The RMS of the considered variations is assigned as the systematic uncertainty related to the unfolding procedure, where the uncertainty is considered as uncorrelated between \z bins. The uncertainties range from 1--10\% and 0.3--7\% for the prompt and non-prompt \jpsi, respectively. 
The reconstruction efficiency of tracks used for jet reconstruction is known with 4\% uncertainty~\cite{ALICE:2019qyj, ALICE:2023jgm, ALICE:2022mur}. Since this efficiency affects the jet-momentum resolution, the corresponding impact of this uncertainty on the measured \jpsi yield in \z intervals is quantified by repeating the unfolding procedure with a new response matrix where 4\% of the tracks are randomly rejected. The systematic uncertainty for a 4\% worse tracking efficiency is considered symmetric, accounting also for a 4\% increased tracking efficiency. The difference between the unfolded distribution using the default and modified response matrix is assigned as the systematic uncertainty. The uncertainties range from 0.1--10\% and 1--11\% for the prompt and non-prompt \jpsi, respectively.

The systematic uncertainties are summarized in Tab.~\ref{tab:systZBins}. The total uncertainties are determined as the sum of the different contributions added in quadrature.

\begin{table}[tb]
\center
\caption{Summary of systematic-uncertainty contributions for the fraction of transverse momentum carried by prompt and non-prompt J/$\psi$ in charged jets. Uncertainties are listed in percentages. For most bins, the highest uncertainties correspond to the lowest \z bin, see text for details.}
  \begin{tabular}{l | c  c }
 \hline
  \multicolumn{1}{c|}{Source} & \multicolumn{2}{c}{Systematic uncertainty [\%]}  \\
    &\it{Prompt} \jpsi &\it{Non-prompt} \jpsi \\
    \hline
Electron selection &\multicolumn{2}{c}{0.7 -- 13.6} \\
Invariant-mass \jpsi PDF &\multicolumn{2}{c}{0.7 -- 8.1} \\
Invariant-mass bkg. PDF &\multicolumn{2}{c}{1.3 -- 3.1} \\
$x$ prompt and non-prompt \jpsi PDFs &\multicolumn{2}{c}{1.3 -- 5.7}      \\
$x$ bkg. PDF &\multicolumn{2}{c}{1.4 -- 11.4}  \\
TRD trigger efficiency &\multicolumn{2}{c}{1.0} \\
Jet unfolding procedure &0.7 -- 10.3 &0.3 -- 6.5 \\
Tracking efficiency &0.1 -- 10.2 &1.4 -- 11.2  \\
    \hline\hline
Total &4.1 -- 25.1 &4.1 -- 23.4  \\
    \hline
  \end{tabular}
  \label{tab:systZBins}
\end{table}

\section{Results}

\begin{figure}[tb]
    \begin{center}
        \includegraphics[width = 0.48\textwidth]{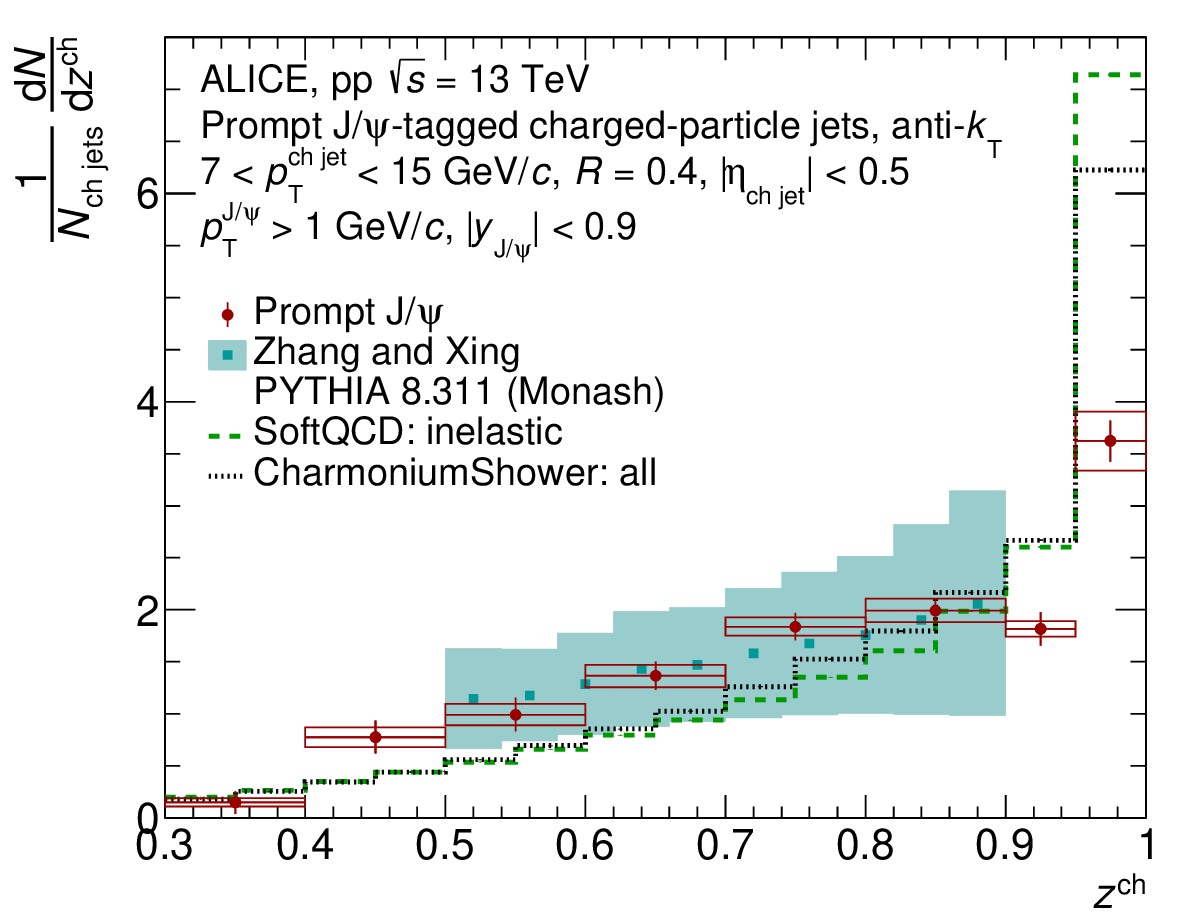}
        \includegraphics[width = 0.48\textwidth]{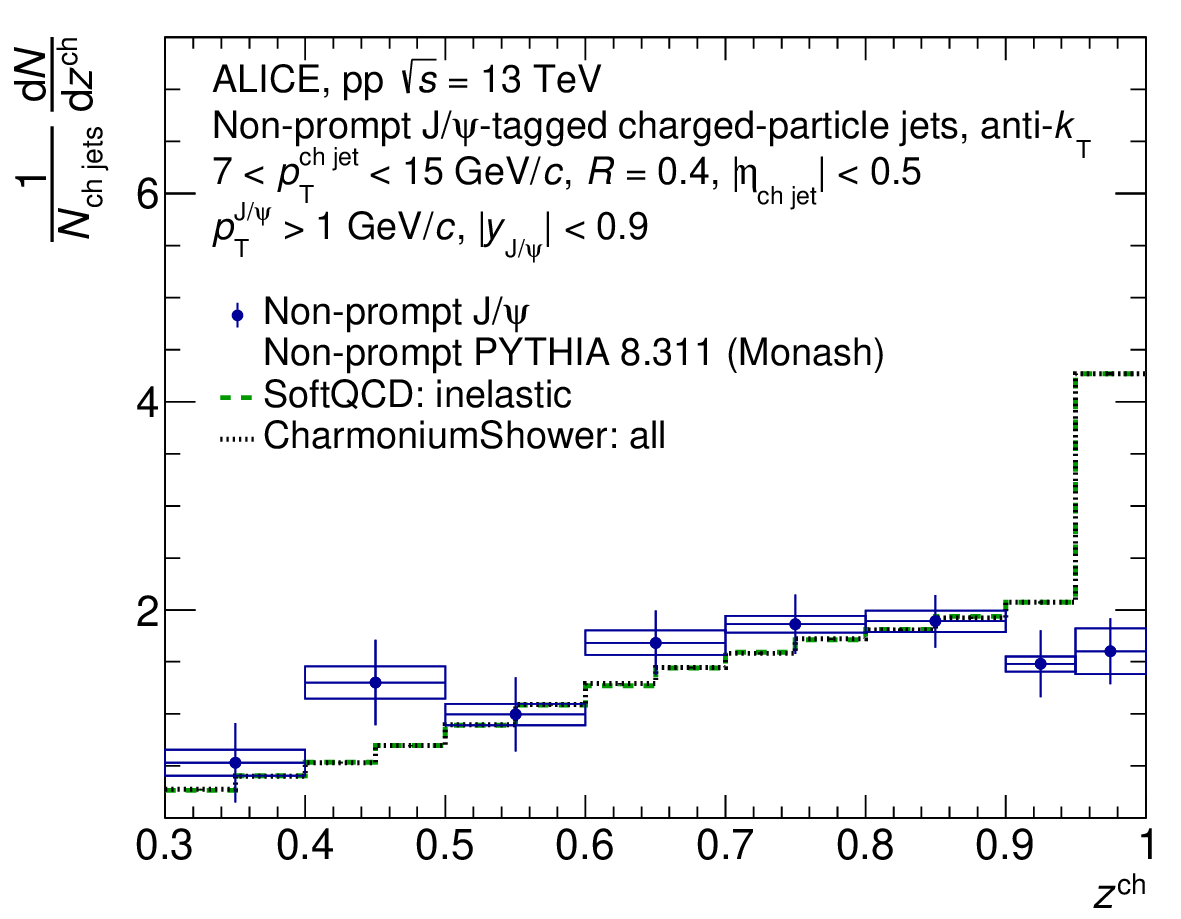}
    \end{center}
    \caption{Fully corrected, normalized prompt (\textit{left}) and non-prompt (\textit{right}) \jpsi transverse-momentum fraction (\z) measured in jets with $7<\ptjetch < 15$~\GeVc for $\ptjpsi > 1$~\GeVc in pp collisions at \thirteen. Measurements are compared to PYTHIA 8.311 calculations obtained using the Monash 2013 tune~\cite{Skands:2014pea}, where \jpsi are produced directly in parton--parton scatterings (black dotted line) and as a part of a parton shower (green dashed line)~\cite{Cooke:2023ldt}. The prompt-\jpsi distribution is compared to a model calculation from Zhang and Xing~\cite{Zhang:2024owr, Kang:2017yde} (cyan box). The error bars represent statistical uncertainties, while the boxes around the data points represent the total systematic uncertainty.}
    \label{fig:results}
\end{figure}

The corrected per-jet \jpsi yield distribution, $1/N_{\rm ch~jets} ({\rm d}N/{\rm d}\z)$, as a function of the fraction of jet transverse momentum carried by the \jpsi, \z, measured for charged-particle jets in the interval $7<\ptjetch < 15$~\GeVc and for $\ptjpsi > 1$~\GeVc, is shown in Fig.~\ref{fig:results} for the prompt (left panel) and non-prompt (right panel) components.
The prompt-\jpsi distribution exhibits an increasing trend for increasing \z, with a clear peak as \z approaches one. The distinct component for $\z \approx 1$ is consistent with isolated \jpsi production, indicating that this is the dominant production mechanism in the low \ptjetch range. 
A qualitatively similar observation was also reported recently by the LHCb collaboration in Ref.~\cite{LHCb:2024ybz} for prompt \psiprime reconstructed in full jets, where the distributions in the low-\ptjet ranges ($8 <\ptjet < 11$~\GeVc and $11 <\ptjet < 15$~\GeVc) show a distinct and growing fraction. With increasing \ptjet, the prompt-\psiprime distributions reported by the LHCb collaboration show a distinct fraction of non-isolated production in addition to the isolated production, where the evolution of the non-isolated component is shown to be driven by the \jpsi-\pt component in each jet-\pt interval.
The distribution corresponding to the non-prompt \jpsi in charged-particle jets, as shown in the right panel of Fig.~\ref{fig:results}, is compatible with a rising trend within the current uncertainties. As non-prompt \jpsi originate from the decay of beauty hadrons, it is expected that the non-prompt \jpsi are produced with a larger fraction of surrounding decay products. 

The prompt and non-prompt \jpsi \z distributions are compared to PYTHIA 8.311 simulations using the Monash 2013 tune~\cite{Skands:2014pea} with two different production settings. In the \texttt{softQCD} production setting, prompt-\jpsi production is determined from fixed-order NRQCD calculations directly in the parton--parton scatterings, while the \texttt{charmoniumShower} setting allows \jpsi production within a parton shower~\cite{Cooke:2023ldt}. The PYTHIA 8 distributions are normalized to unity in the range $0.3 < \z \leq 1$, similarly to data.
For the prompt \jpsi, the PYTHIA 8 simulations with both settings qualitatively reproduce the rising trend observed in data, however, the \z distribution is significantly above the measurement in the highest \z interval, indicating that the fraction of isolated \jpsi is overestimated. A similar trend of the simulation results is observed also for the non-prompt \jpsi, with a good compatibility to data up to $\z = 0.9$. For $\z > 0.9$, the PYTHIA 8 results exhibit a pronounced peak, not observed in data.
Previous measurements of the non-prompt charmonium are well reproduced by PYTHIA 8 simulations~\cite{LHCb:2017llq, LHCb:2024ybz}, where the distributions are peaked around intermediate $z$ values. As non-prompt \jpsi originate from the decay of beauty hadrons, it is expected that the \jpsi is accompanied by other decay products and will thus, cluster into a jet. This would then lead to a larger fraction of non-isolated \jpsi production, as seen in data. 
The measurements presented by the LHCb collaboration cover a different phase-space region, as both the jet radius and \ptjet are larger than for the measurement presented here. With increasing jet radius, one would expect less isolated-\jpsi production, as more surrounding particles, including decay products, are included in the jet, shifting the distribution to lower $z$ values. 
However, the possible tension between data and simulations in the highest \z bins may reflect the challenge of simulating hadronization of low-momentum jets. 

The left panel of Fig.~\ref{fig:results} shows the prompt \jpsi \z distribution compared to a model calculation from Zhang and Xing~\cite{Zhang:2024owr, Kang:2017yde}. In this calculation, the differential cross section of a parton distribution in a jet is simulated using MadGraph~\cite{Alwall:2014hca} for the creation of the hard parton and using PYTHIA 8 for the parton shower down to the scale of two times the charm mass. 
The short-distance coefficients, which embody the perturbative part of the calculation, are taken from Ma et al.~\cite{Ma:2013yla}, while the non-perturbative part is represented by long-distance matrix elements (LDMEs) taken from Bodwin et al.~\cite{Bodwin:2014gia}. These calculations cover the range $0.5 < \z < 0.9$ and in order to compare them to the data, they are normalized such that they match the integral of the measurement in the range $0.5 < \z < 0.9$. A very good compatibility with the data is observed, although the theoretical calculations have very large uncertainties, which come from the uncertainty on the LDMEs.

\begin{figure}[tb]
    \begin{center}
    \includegraphics[width = 0.6\textwidth]{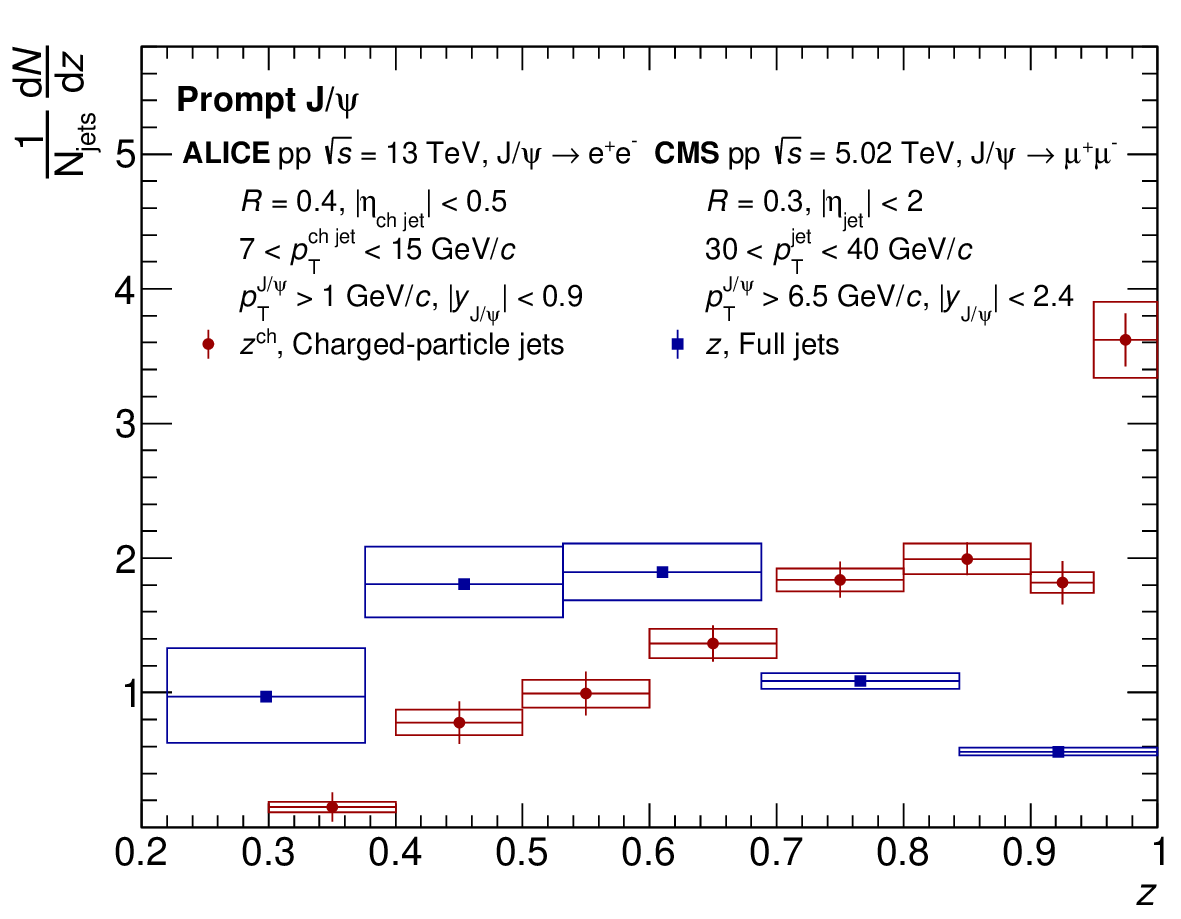}
    \end{center}
    \caption{Comparison of the measured prompt \jpsi transverse-momentum fraction (\z) measured in charged-particle jets with $7<\ptjetch < 15$~\GeVc for $\ptjpsi > 1$~\GeVc in pp collisions at \thirteen (red point) with the prompt-\jpsi transverse-momentum fraction ($z$) measured in full jets with $30<\ptjet < 40$~\GeVc for $\ptjpsi > 6.5$~\GeVc in pp collisions at \five reported by CMS (blue box)~\cite{CMS:2021puf}. The error bars represent statistical uncertainties, while the boxes around the data points represent the total systematic uncertainty.}
    \label{fig:resultsCMS}
\end{figure}

Figure~\ref{fig:resultsCMS} shows the distribution of the fraction of transverse momentum carried by prompt \jpsi in charged-particle jets with $7 < \ptjetch < 15$~\GeVc, $|\eta_{\rm jet}| < 0.5$, and $R = 0.4$ obtained in this analysis in pp collisions at \thirteen compared to the results obtained using full jets with $30 < \ptjet < 40$~\GeVc, $|\eta_{\rm jet}| < 2$, and $R = 0.3$ in pp collisions at \five reported by CMS in Ref.~\cite{CMS:2021puf}. The distribution measured by CMS peaks at a $z$ value in the range $0.5-0.6$, and does not exhibit an isolated-\jpsi peak at large $z$. 
The difference between the ALICE and CMS data seen in Fig~\ref{fig:resultsCMS}  is expected to be, in the first order, caused by the different kinematic regions. The recent measurement of prompt \psiprime reported by the LHCb collaboration~\cite{LHCb:2024ybz} supports this, where the transverse-momentum fraction evolves with jet \pt from a shape resembling the prompt-\jpsi ALICE measurement to one more similar to the reported CMS results. In addition, the CMS measurement uses full jets, while in the ALICE measurement charged-particle jets were used, meaning that approximately one third of the jet energy is missing in the case of the charged-particle jets, leading to \z distributions shifted towards higher $z$.

\section{Summary}
This paper presents results on the production of prompt and non-prompt \jpsi mesons in charged-particle jets using data from pp collisions at \thirteen recorded by the ALICE detector at the CERN LHC. The data sample is triggered based on information from the Transition Radiation Detector. The distribution of the transverse-momentum fraction for prompt and non-prompt \jpsi in charged-particle jets is reported for jets with $7 < \ptjetch < 15$~\GeVc, $|\eta_{\text{jet}}|< 0.5$, and resolution parameter $R=0.4$. The \jpsi candidates are reconstructed at midrapidity through the electron--positron decay channel for \jpsi with $\ptjpsi > 1$~\GeVc. 

The presented measurement is provided in a unique kinematic range at the LHC, investigating for the first time \jpsi production in charged-particle jets at very low jet \pt, sensitive to potential factorization-breaking effects. Both the ALICE results on the prompt and non-prompt \jpsi production within charged-particle jets are found to be qualitatively well reproduced by PYTHIA 8.311 simulations for $\z < 0.9$. As \z approaches one, the simulations overshoot the measured data, which indicates an overestimation of the fraction of isolated \jpsi. The possible tension may also reflect the challenge of simulating hadronization of low-momentum jets. 
The trend of the prompt-\jpsi distribution can also be described by a pQCD calculation for charmonium fragmentation in the range $0.5 < \z < 0.9$. 
Although the measurements are from different kinematic regions, it is noteworthy that the LHCb results for prompt $\psi(2S)$ are in qualitative agreement with the prompt-\jpsi measurements in the same jet-\pt range~\cite{LHCb:2024ybz}, and gradually evolve with jet \pt to a shape resembling the reported CMS results of prompt \jpsi at significantly higher jet \pt. 

By exploiting the much larger datasets collected by ALICE during the LHC Run 3 and 4, more differential results are expected in the near future. This will give the possibility of further investigating the kinematic dependencies of the \jpsi \z distributions. Foreseen measurements of the prompt and non-prompt \jpsi production in jets in heavy-ion collisions are expected to provide further information on the energy-loss mechanisms of heavy quarks in the medium.


\newenvironment{acknowledgement}{\relax}{\relax}
\begin{acknowledgement}
\section*{Acknowledgements}

The ALICE Collaboration would like to thank all its engineers and technicians for their invaluable contributions to the construction of the experiment and the CERN accelerator teams for the outstanding performance of the LHC complex.
The ALICE Collaboration gratefully acknowledges the resources and support provided by all Grid centres and the Worldwide LHC Computing Grid (WLCG) collaboration.
The ALICE Collaboration acknowledges the following funding agencies for their support in building and running the ALICE detector:
A. I. Alikhanyan National Science Laboratory (Yerevan Physics Institute) Foundation (ANSL), State Committee of Science and World Federation of Scientists (WFS), Armenia;
Austrian Academy of Sciences, Austrian Science Fund (FWF): [M 2467-N36] and Nationalstiftung f\"{u}r Forschung, Technologie und Entwicklung, Austria;
Ministry of Communications and High Technologies, National Nuclear Research Center, Azerbaijan;
Rede Nacional de Física de Altas Energias (Renafae), Financiadora de Estudos e Projetos (Finep), Funda\c{c}\~{a}o de Amparo \`{a} Pesquisa do Estado de S\~{a}o Paulo (FAPESP) and The Sao Paulo Research Foundation  (FAPESP), Brazil;
Bulgarian Ministry of Education and Science, within the National Roadmap for Research Infrastructures 2020-2027 (object CERN), Bulgaria;
Ministry of Education of China (MOEC) , Ministry of Science \& Technology of China (MSTC) and National Natural Science Foundation of China (NSFC), China;
Ministry of Science and Education and Croatian Science Foundation, Croatia;
Centro de Aplicaciones Tecnol\'{o}gicas y Desarrollo Nuclear (CEADEN), Cubaenerg\'{\i}a, Cuba;
Ministry of Education, Youth and Sports of the Czech Republic, Czech Republic;
The Danish Council for Independent Research | Natural Sciences, the VILLUM FONDEN and Danish National Research Foundation (DNRF), Denmark;
Helsinki Institute of Physics (HIP), Finland;
Commissariat \`{a} l'Energie Atomique (CEA) and Institut National de Physique Nucl\'{e}aire et de Physique des Particules (IN2P3) and Centre National de la Recherche Scientifique (CNRS), France;
Bundesministerium f\"{u}r Forschung, Technologie und Raumfahrt (BMFTR) and GSI Helmholtzzentrum f\"{u}r Schwerionenforschung GmbH, Germany;
National Research, Development and Innovation Office, Hungary;
Department of Atomic Energy Government of India (DAE), Department of Science and Technology, Government of India (DST), University Grants Commission, Government of India (UGC) and Council of Scientific and Industrial Research (CSIR), India;
National Research and Innovation Agency - BRIN, Indonesia;
Istituto Nazionale di Fisica Nucleare (INFN), Italy;
Japanese Ministry of Education, Culture, Sports, Science and Technology (MEXT) and Japan Society for the Promotion of Science (JSPS) KAKENHI, Japan;
Consejo Nacional de Ciencia (CONACYT) y Tecnolog\'{i}a, through Fondo de Cooperaci\'{o}n Internacional en Ciencia y Tecnolog\'{i}a (FONCICYT) and Direcci\'{o}n General de Asuntos del Personal Academico (DGAPA), Mexico;
Nederlandse Organisatie voor Wetenschappelijk Onderzoek (NWO), Netherlands;
The Research Council of Norway, Norway;
Pontificia Universidad Cat\'{o}lica del Per\'{u}, Peru;
Ministry of Science and Higher Education, National Science Centre and WUT ID-UB, Poland;
Korea Institute of Science and Technology Information and National Research Foundation of Korea (NRF), Republic of Korea;
Ministry of Education and Scientific Research, Institute of Atomic Physics, Ministry of Research and Innovation and Institute of Atomic Physics and Universitatea Nationala de Stiinta si Tehnologie Politehnica Bucuresti, Romania;
Ministerstvo skolstva, vyskumu, vyvoja a mladeze SR, Slovakia;
National Research Foundation of South Africa, South Africa;
Swedish Research Council (VR) and Knut \& Alice Wallenberg Foundation (KAW), Sweden;
European Organization for Nuclear Research, Switzerland;
Suranaree University of Technology (SUT), National Science and Technology Development Agency (NSTDA) and National Science, Research and Innovation Fund (NSRF via PMU-B B05F650021), Thailand;
Turkish Energy, Nuclear and Mineral Research Agency (TENMAK), Turkey;
National Academy of  Sciences of Ukraine, Ukraine;
Science and Technology Facilities Council (STFC), United Kingdom;
National Science Foundation of the United States of America (NSF) and United States Department of Energy, Office of Nuclear Physics (DOE NP), United States of America.
In addition, individual groups or members have received support from:
FORTE project, reg.\ no.\ CZ.02.01.01/00/22\_008/0004632, Czech Republic, co-funded by the European Union, Czech Republic;
European Research Council (grant no. 950692), European Union;
Deutsche Forschungs Gemeinschaft (DFG, German Research Foundation) ``Neutrinos and Dark Matter in Astro- and Particle Physics'' (grant no. SFB 1258), Germany;
FAIR - Future Artificial Intelligence Research, funded by the NextGenerationEU program (Italy).

\end{acknowledgement}

\bibliographystyle{utphys}   
\bibliography{bibliography}

\newpage
\appendix

%
%

\section{The ALICE Collaboration}
\label{app:collab}
\begin{flushleft} 
\small

D.A.H.~Abdallah\,\orcidlink{0000-0003-4768-2718}\,$^{\rm 134}$, 
I.J.~Abualrob\,\orcidlink{0009-0005-3519-5631}\,$^{\rm 112}$, 
S.~Acharya\,\orcidlink{0000-0002-9213-5329}\,$^{\rm 49}$, 
G.~Aglieri Rinella\,\orcidlink{0000-0002-9611-3696}\,$^{\rm 32}$, 
L.~Aglietta\,\orcidlink{0009-0003-0763-6802}\,$^{\rm 24}$, 
N.~Agrawal\,\orcidlink{0000-0003-0348-9836}\,$^{\rm 25}$, 
Z.~Ahammed\,\orcidlink{0000-0001-5241-7412}\,$^{\rm 132}$, 
S.~Ahmad\,\orcidlink{0000-0003-0497-5705}\,$^{\rm 15}$, 
I.~Ahuja\,\orcidlink{0000-0002-4417-1392}\,$^{\rm 36}$, 
Z.~Akbar$^{\rm 79}$
% ...

\section*{Affiliation Notes}

$^{\rm I}$ Deceased\\
$^{\rm II}$ Also at: Max-Planck-Institut fur Physik, Munich, Germany\\
$^{\rm III}$ Also at: Czech Technical University in Prague (CZ)\\
$^{\rm IV}$ Also at: Instituto de Fisica da Universidade de Sao Paulo\\
$^{\rm V}$ Also at: Dipartimento DET del Politecnico di Torino, Turin, Italy\\
$^{\rm VI}$ Also at: Department of Applied Physics, Aligarh Muslim University, Aligarh, India\\
$^{\rm VII}$ Also at: Institute of Theoretical Physics, University of Wroclaw, Poland\\
$^{\rm VIII}$ Also at: Facultad de Ciencias, Universidad Nacional Aut\'{o}noma de M\'{e}xico, Mexico City, Mexico\\

\section*{Collaboration Institutes}

$^{1}$ A.I. Alikhanyan National Science Laboratory (Yerevan Physics Institute) Foundation, Yerevan, Armenia\\
$^{2}$ AGH University of Krakow, Cracow, Poland\\
$^{3}$ Bogolyubov Institute for Theoretical Physics, National Academy of Sciences of Ukraine, Kyiv, Ukraine\\
$^{4}$ Bose Institute, Department of Physics  and Centre for Astroparticle Physics and Space Science (CAPSS), Kolkata, India\\
$^{5}$ California Polytechnic State University, San Luis Obispo, California, United States\\
$^{6}$ Central China Normal University, Wuhan, China\\
$^{7}$ Centro de Aplicaciones Tecnol\'{o}gicas y Desarrollo Nuclear (CEADEN), Havana, Cuba\\
$^{8}$ Centro de Investigaci\'{o}n y de Estudios Avanzados (CINVESTAV), Mexico City and M\'{e}rida, Mexico\\
$^{9}$ Chicago State University, Chicago, Illinois, United States\\
$^{10}$ China Nuclear Data Center, China Institute of Atomic Energy, Beijing, China\\
$^{11}$ China University of Geosciences, Wuhan, China\\
$^{12}$ Chungbuk National University, Cheongju, Republic of Korea\\
$^{13}$ Comenius University Bratislava, Faculty of Mathematics, Physics and Informatics, Bratislava, Slovak Republic\\
$^{14}$ Creighton University, Omaha, Nebraska, United States\\
$^{15}$ Department of Physics, Aligarh Muslim University, Aligarh, India\\
$^{16}$ Department of Physics, Pusan National University, Pusan, Republic of Korea\\
$^{17}$ Department of Physics, Sejong University, Seoul, Republic of Korea\\
$^{18}$ Department of Physics, University of California, Berkeley, California, United States\\
$^{19}$ Department of Physics, University of Oslo, Oslo, Norway\\
$^{20}$ Department of Physics and Technology, University of Bergen, Bergen, Norway\\
$^{21}$ Dipartimento di Fisica, Universit\`{a} di Pavia, Pavia, Italy\\
$^{22}$ Dipartimento di Fisica dell'Universit\`{a} and Sezione INFN, Cagliari, Italy\\
$^{23}$ Dipartimento di Fisica dell'Universit\`{a} and Sezione INFN, Trieste, Italy\\
$^{24}$ Dipartimento di Fisica dell'Universit\`{a} and Sezione INFN, Turin, Italy\\
$^{25}$ Dipartimento di Fisica e Astronomia dell'Universit\`{a} and Sezione INFN, Bologna, Italy\\
$^{26}$ Dipartimento di Fisica e Astronomia dell'Universit\`{a} and Sezione INFN, Catania, Italy\\
$^{27}$ Dipartimento di Fisica e Astronomia dell'Universit\`{a} and Sezione INFN, Padova, Italy\\
$^{28}$ Dipartimento di Fisica `E.R.~Caianiello' dell'Universit\`{a} and Gruppo Collegato INFN, Salerno, Italy\\
$^{29}$ Dipartimento DISAT del Politecnico and Sezione INFN, Turin, Italy\\
$^{30}$ Dipartimento di Scienze MIFT, Universit\`{a} di Messina, Messina, Italy\\
$^{31}$ Dipartimento Interateneo di Fisica `M.~Merlin' and Sezione INFN, Bari, Italy\\
$^{32}$ European Organization for Nuclear Research (CERN), Geneva, Switzerland\\
$^{33}$ Faculty of Electrical Engineering, Mechanical Engineering and Naval Architecture, University of Split, Split, Croatia\\
$^{34}$ Faculty of Nuclear Sciences and Physical Engineering, Czech Technical University in Prague, Prague, Czech Republic\\
$^{35}$ Faculty of Physics, Sofia University, Sofia, Bulgaria\\
$^{36}$ Faculty of Science, P.J.~\v{S}af\'{a}rik University, Ko\v{s}ice, Slovak Republic\\
$^{37}$ Faculty of Technology, Environmental and Social Sciences, Bergen, Norway\\
$^{38}$ Frankfurt Institute for Advanced Studies, Johann Wolfgang Goethe-Universit\"{a}t Frankfurt, Frankfurt, Germany\\
$^{39}$ Fudan University, Shanghai, China\\
$^{40}$ Gauhati University, Department of Physics, Guwahati, India\\
$^{41}$ Helmholtz-Institut f\"{u}r Strahlen- und Kernphysik, Rheinische Friedrich-Wilhelms-Universit\"{a}t Bonn, Bonn, Germany\\
$^{42}$ Helsinki Institute of Physics (HIP), Helsinki, Finland\\
$^{43}$ High Energy Physics Group,  Universidad Aut\'{o}noma de Puebla, Puebla, Mexico\\
$^{44}$ Horia Hulubei National Institute of Physics and Nuclear Engineering, Bucharest, Romania\\
$^{45}$ HUN-REN Wigner Research Centre for Physics, Budapest, Hungary\\
$^{46}$ Indian Institute of Technology Bombay (IIT), Mumbai, India\\
$^{47}$ Indian Institute of Technology Indore, Indore, India\\
$^{48}$ INFN, Laboratori Nazionali di Frascati, Frascati, Italy\\
$^{49}$ INFN, Sezione di Bari, Bari, Italy\\
$^{50}$ INFN, Sezione di Bologna, Bologna, Italy\\
$^{51}$ INFN, Sezione di Cagliari, Cagliari, Italy\\
$^{52}$ INFN, Sezione di Catania, Catania, Italy\\
$^{53}$ INFN, Sezione di Padova, Padova, Italy\\
$^{54}$ INFN, Sezione di Pavia, Pavia, Italy\\
$^{55}$ INFN, Sezione di Torino, Turin, Italy\\
$^{56}$ INFN, Sezione di Trieste, Trieste, Italy\\
$^{57}$ Inha University, Incheon, Republic of Korea\\
$^{58}$ Institute for Gravitational and Subatomic Physics (GRASP), Utrecht University/Nikhef, Utrecht, Netherlands\\
$^{59}$ Institute of Experimental Physics, Slovak Academy of Sciences, Ko\v{s}ice, Slovak Republic\\
$^{60}$ Institute of Physics, Homi Bhabha National Institute, Bhubaneswar, India\\
$^{61}$ Institute of Physics of the Czech Academy of Sciences, Prague, Czech Republic\\
$^{62}$ Institute of Space Science (ISS), Bucharest, Romania\\
$^{63}$ Institut f\"{u}r Kernphysik, Johann Wolfgang Goethe-Universit\"{a}t Frankfurt, Frankfurt, Germany\\
$^{64}$ Instituto de Ciencias Nucleares, Universidad Nacional Aut\'{o}noma de M\'{e}xico, Mexico City, Mexico\\
$^{65}$ Instituto de F\'{i}sica, Universidade Federal do Rio Grande do Sul (UFRGS), Porto Alegre, Brazil\\
$^{66}$ Instituto de F\'{\i}sica, Universidad Nacional Aut\'{o}noma de M\'{e}xico, Mexico City, Mexico\\
$^{67}$ iThemba LABS, National Research Foundation, Somerset West, South Africa\\
$^{68}$ Jeonbuk National University, Jeonju, Republic of Korea\\
$^{69}$ Korea Institute of Science and Technology Information, Daejeon, Republic of Korea\\
$^{70}$ Laboratoire de Physique Subatomique et de Cosmologie, Universit\'{e} Grenoble-Alpes, CNRS-IN2P3, Grenoble, France\\
$^{71}$ Lawrence Berkeley National Laboratory, Berkeley, California, United States\\
$^{72}$ Lund University Department of Physics, Division of Particle Physics, Lund, Sweden\\
$^{73}$ Marietta Blau Institute, Vienna, Austria\\
$^{74}$ Nagasaki Institute of Applied Science, Nagasaki, Japan\\
$^{75}$ Nara Women{'}s University (NWU), Nara, Japan\\
$^{76}$ National Centre for Nuclear Research, Warsaw, Poland\\
$^{77}$ National Institute of Science Education and Research, Homi Bhabha National Institute, Jatni, India\\
$^{78}$ National Nuclear Research Center, Baku, Azerbaijan\\
$^{79}$ National Research and Innovation Agency - BRIN, Jakarta, Indonesia\\
$^{80}$ Niels Bohr Institute, University of Copenhagen, Copenhagen, Denmark\\
$^{81}$ Nikhef, National institute for subatomic physics, Amsterdam, Netherlands\\
$^{82}$ Nuclear Physics Group, STFC Daresbury Laboratory, Daresbury, United Kingdom\\
$^{83}$ Nuclear Physics Institute of the Czech Academy of Sciences, Husinec-\v{R}e\v{z}, Czech Republic\\
$^{84}$ Oak Ridge National Laboratory, Oak Ridge, Tennessee, United States\\
$^{85}$ Ohio State University, Columbus, Ohio, United States\\
$^{86}$ Physics department, Faculty of science, University of Zagreb, Zagreb, Croatia\\
$^{87}$ Physics Department, Panjab University, Chandigarh, India\\
$^{88}$ Physics Department, University of Jammu, Jammu, India\\
$^{89}$ Physics Program and International Institute for Sustainability with Knotted Chiral Meta Matter (WPI-SKCM$^{2}$), Hiroshima University, Hiroshima, Japan\\
$^{90}$ Physikalisches Institut, Eberhard-Karls-Universit\"{a}t T\"{u}bingen, T\"{u}bingen, Germany\\
$^{91}$ Physikalisches Institut, Ruprecht-Karls-Universit\"{a}t Heidelberg, Heidelberg, Germany\\
$^{92}$ Physik Department, Technische Universit\"{a}t M\"{u}nchen, Munich, Germany\\
$^{93}$ Politecnico di Bari and Sezione INFN, Bari, Italy\\
$^{94}$ Research Division and ExtreMe Matter Institute EMMI, GSI Helmholtzzentrum f\"ur Schwerionenforschung GmbH, Darmstadt, Germany\\
$^{95}$ Saga University, Saga, Japan\\
$^{96}$ Saha Institute of Nuclear Physics, Homi Bhabha National Institute, Kolkata, India\\
$^{97}$ School of Physics and Astronomy, University of Birmingham, Birmingham, United Kingdom\\
$^{98}$ Secci\'{o}n F\'{\i}sica, Departamento de Ciencias, Pontificia Universidad Cat\'{o}lica del Per\'{u}, Lima, Peru\\
$^{99}$ SUBATECH, IMT Atlantique, Nantes Universit\'{e}, CNRS-IN2P3, Nantes, France\\
$^{100}$ Sungkyunkwan University, Suwon City, Republic of Korea\\
$^{101}$ Suranaree University of Technology, Nakhon Ratchasima, Thailand\\
$^{102}$ Technical University of Ko\v{s}ice, Ko\v{s}ice, Slovak Republic\\
$^{103}$ The Henryk Niewodniczanski Institute of Nuclear Physics, Polish Academy of Sciences, Cracow, Poland\\
$^{104}$ The University of Texas at Austin, Austin, Texas, United States\\
$^{105}$ Universidad Aut\'{o}noma de Sinaloa, Culiac\'{a}n, Mexico\\
$^{106}$ Universidade de S\~{a}o Paulo (USP), S\~{a}o Paulo, Brazil\\
$^{107}$ Universidade Estadual de Campinas (UNICAMP), Campinas, Brazil\\
$^{108}$ Universidade Federal do ABC, Santo Andre, Brazil\\
$^{109}$ Universitatea Nationala de Stiinta si Tehnologie Politehnica Bucuresti, Bucharest, Romania\\
$^{110}$ University of Cape Town, Cape Town, South Africa\\
$^{111}$ University of Derby, Derby, United Kingdom\\
$^{112}$ University of Houston, Houston, Texas, United States\\
$^{113}$ University of Jyv\"{a}skyl\"{a}, Jyv\"{a}skyl\"{a}, Finland\\
$^{114}$ University of Kansas, Lawrence, Kansas, United States\\
$^{115}$ University of Liverpool, Liverpool, United Kingdom\\
$^{116}$ University of Science and Technology of China, Hefei, China\\
$^{117}$ University of Silesia in Katowice, Katowice, Poland\\
$^{118}$ University of South-Eastern Norway, Kongsberg, Norway\\
$^{119}$ University of Tennessee, Knoxville, Tennessee, United States\\
$^{120}$ University of the Witwatersrand, Johannesburg, South Africa\\
$^{121}$ University of Tokyo, Tokyo, Japan\\
$^{122}$ University of Tsukuba, Tsukuba, Japan\\
$^{123}$ Universit\"{a}t M\"{u}nster, Institut f\"{u}r Kernphysik, M\"{u}nster, Germany\\
$^{124}$ Universit\'{e} Clermont Auvergne, CNRS/IN2P3, LPC, Clermont-Ferrand, France\\
$^{125}$ Universit\'{e} de Lyon, CNRS/IN2P3, Institut de Physique des 2 Infinis de Lyon, Lyon, France\\
$^{126}$ Universit\'{e} de Strasbourg, CNRS, IPHC UMR 7178, F-67000 Strasbourg, France, Strasbourg, France\\
$^{127}$ Universit\'{e} Paris-Saclay, Centre d'Etudes de Saclay (CEA), IRFU, D\'{e}partment de Physique Nucl\'{e}aire (DPhN), Saclay, France\\
$^{128}$ Universit\'{e}  Paris-Saclay, CNRS/IN2P3, IJCLab, Orsay, France\\
$^{129}$ Universit\`{a} degli Studi di Foggia, Foggia, Italy\\
$^{130}$ Universit\`{a} del Piemonte Orientale, Vercelli, Italy\\
$^{131}$ Universit\`{a} di Brescia, Brescia, Italy\\
$^{132}$ Variable Energy Cyclotron Centre, Homi Bhabha National Institute, Kolkata, India\\
$^{133}$ Warsaw University of Technology, Warsaw, Poland\\
$^{134}$ Wayne State University, Detroit, Michigan, United States\\
$^{135}$ Yale University, New Haven, Connecticut, United States\\
$^{136}$ Yildiz Technical University, Istanbul, Turkey\\
$^{137}$ Yonsei University, Seoul, Republic of Korea\\
$^{138}$ Affiliated with an institute formerly covered by a cooperation agreement with CERN\\
$^{139}$ Affiliated with an international laboratory covered by a cooperation agreement with CERN.\\

\end{flushleft} 
  
\end{document}